\documentclass{article}
\usepackage{amsmath}
\usepackage{graphicx}
\usepackage{amsfonts}
\usepackage{amssymb}
\usepackage{geometry}

\newcommand{\bee}{\begin{equation}}
\newcommand{\eend}{\end{equation}}

\newcommand{\bea}{\begin{eqnarray}}
\newcommand{\eea}{\end{eqnarray}}

\begin{document}

\title{Note on "Finite Field-Energy and Interparticle Potential in
Logarithmic Electrodynamics"}
\author{ Dmitry M. Gitman$^{1,2,3}$\thanks{%
Electronic address: gitman@dfn.if.usp.br} and Anatoly E. Shabad$^{2,3}$%
\thanks{%
Electronic address: shabad@lpi.ru} \\
%EndAName
$^{1}$\textsl{Instituto de F\'{\i}sica, Universidade de S\~{a}o Paulo, }\\
\textsl{Caixa Postal 66318, CEP 05508-090, S\~{a}o Paulo, S. P., Brazil} \\
$^{2}$\textsl{P. N. Lebedev Physics Institute, Leninsky Prospekt 53, Moscow
117924, Russia} \\
$^{3}$\textsl{P. Tomsk State University, Lenin Prospekt 36, Tomsk 634050,
Russia} }
\maketitle

\begin{abstract}
We propose an identification of the free parameter in the model of nonlinear
electrodynamics proposed in \cite{GaeHel} by equating the second term in the
power expansion of its Lagrangian with that in the expansion of the
Heiseberg-Euler Lagrangian. The resulting value of the field-energy of a
point-like charge makes 0.988 of the electron mass, if the charge is that of
the electron.
\end{abstract}

\bigskip Recently Gaete and Helay$\mathrm{{\ddot{e}}}$l-Neto \cite{GaeHel}
proposed a nonlinear Lagrangian for what they called the logarithmic
electrodynamics, which, in the special case of presence of only electric
field $E$, has the form
\begin{equation}
L=-\beta ^{2}\ln \left[ 1-\frac{E^{2}}{2\beta ^{2}}\right] ,  \label{L}
\end{equation}%
where $\ \beta $ is a nonlinear coupling constant. For small fields one has $%
L=\frac{1}{2}E^{2},$ in correspondence with the classical electrodynamics.
The same as in the Born-Infeld theory this Lagrangian has a singularity at a
certain value of the electric field, namely $E=2^{1/2}\beta ,$ which makes
the maximum value of the electric field produced by a point-like charge $Q$%
\begin{equation}
E(r)=\frac{4\pi \beta ^{2}}{Q}\frac{1}{\left( r^{4}+\frac{2Q^{2}}{\beta ^{2}}%
\right) ^{1/2}-r^{2}}  \label{Electric}
\end{equation}%
in the origin of coordinates $r=0$, where the charge is placed. The fineness
of the maximum electric field results in convergence of the integral for the
electrostatic field-energy produced by the charge, again the same as in the
Born-Infeld model. It is calculated in \cite{GaeHel} as the space integral
over the Noether energy density
\begin{equation*}
\Theta ^{00}=\frac{E^{2}}{1-\frac{E^{2}}{2\beta ^{2}}}+\beta ^{2}\ln \left(
1-\frac{E^{2}}{2\beta ^{2}}\right)
\end{equation*}%
taken on the field (\ref{Electric}) to be%
\begin{equation}
U=4\pi \int_{0}^{\infty }\Theta ^{00}r^{2}dr=0.391\sqrt{e^{3}\beta }.
\label{U}
\end{equation}

\bigskip To fix the model, the authors of Ref.\cite{GaeHel} have proposed,
when setting the point charge equal to that of the electron, $Q=e,$ to
equalize the maximum field $E=2^{1/2}\beta $ with the characteristic QED
value $m^{2}/e,$ where $m$ is the electron mass. This led them to the value
of the field-energy $U=4\pi \int_{0}^{1/m}\Theta ^{00}r^{2}dr=8.67\times
10^{-4}m$ that makes a small part of the electron mass. In contrast to this
suggestion, ours is to equalize the Lagrangian (\ref{L}) and the
Heisenberg-Euler Lagrangian within the accuracy to the second power in the
field strength squared \cite{BerLifPit} (similarly to what we \cite%
{CosGitSha} and, previously, other authors (\textit{e.g., } \cite{Davila})
acted in different models following \cite{Infeld}) $\frac{E^{2}}{2}+\frac{%
E^{4}}{8\beta ^{2}}=\frac{E^{2}}{2}+\frac{E^{4}}{8}\frac{e^{4}}{45\pi
^{2}m^{4}},$ which supplies the value
\begin{equation}
\beta =\frac{m^{2}}{e}\frac{3\pi \sqrt{5}}{e}  \label{beta}
\end{equation}%
to the nonlinear coupling constant $\beta $ in (\ref{L}) \ and the maximum
electric field $3\pi \sqrt{10}/e$ $=98$ times the characteristic value $%
m^{2}/e$ . \ With the value \ref{beta} and taking into account that $%
e^{2}/4\pi =1/137$, the field energy of the point charge (\ref{U}) becomes
impressively close to the electron mass:%
\begin{equation}
U=0.988m.  \label{0.988}
\end{equation}

\bigskip If one imagined that the whole mass of the electron might have been
of purely electrostatic origin, $U=m,$\ such assumption would imply via (\ref%
{U}) and (\ref{beta}) the equality%
\begin{equation*}
0.391\sqrt{e3\pi \sqrt{5}}=1,
\end{equation*}%
from which the value $1/130$ would follow for $e^{2}/4\pi .$ Certainly,
other interactions, in which electron is involved, must also contribute into
its field-mass.. The first of them all is the contribution of the static
magnetic dipole field of the electron. Eq. (\ref{0.988}) leaves yet the
spare room of 1.2\% for such extra contributions.

Supported by FAPESP under Processo 2014/08970-1, by RFFI under Project
14-02-01171 and by TSU via Competitiveness Improvement Program.

\end{document}